\documentclass[aps, prd, twocolumn, nofootinbib, noshowpacs, preprintnumbers, longbibliography, floatfix]{revtex4-2}
\usepackage[utf8]{inputenc}
\usepackage[T1]{fontenc}
\usepackage{lmodern}
\usepackage{indentfirst}
\usepackage{graphicx}
\usepackage{rotating}
\usepackage{placeins}
\usepackage{float}
\usepackage{amsmath, amsthm, amssymb, amsfonts}
\usepackage{mathtools}
\usepackage{adjustbox}
\usepackage{diagbox}
\usepackage{array}
\usepackage{multirow}
\usepackage[ruled,vlined]{algorithm2e}
\usepackage{xcolor}
\usepackage{physics}
\usepackage{tikz}
\usetikzlibrary{quantikz2}
\usepackage{subfigure}
\usepackage{braket}
\usepackage{slashed}
\usepackage[colorlinks,
    citecolor={blue!70!black},
    urlcolor={blue!70!black},
    linkcolor={red!70!black},
    hyperindex,breaklinks]{hyperref}
\usepackage{orcidlink}

\usepackage{soul}

\newcommand{\FirstAffiliation}{\affiliation{
    Universit\'{e} Paris-Saclay,
    CEA,
    List,
    F-91120,
    Palaiseau,
    France
    }}
    
\begin{document}
%
%
%

%%%%% TITLE

\title{Quantum Error Mitigation by Global Randomized Error Cancellation for Adiabatic Evolution in the Schwinger Model}

%%%%% AUTHORS

% \thanks{corresponding author}

\author{Oleg Kaikov \orcidlink{0000-0002-9473-7294}}
\email{oleg.kaikov@cea.fr}
\FirstAffiliation

\author{Theo Saporiti \orcidlink{0009-0008-9738-3402}}
\FirstAffiliation

\author{Vasily Sazonov \orcidlink{0000-0002-8152-0221}}
\FirstAffiliation

\author{Mohamed Tamaazousti \orcidlink{0000-0002-3947-9069}}
\FirstAffiliation

%%%%% DATE

\date{\today}

%%%%% ABSTRACT

\begin{abstract}
\noindent We extend the global randomized error cancellation (GREC) method for quantum error mitigation (QEM) in an application to adiabatic evolution of states on a noisy quantum device. We apply the adiabatic GREC method to the evolution of eigenstates in the lattice Schwinger model on a simulated quantum device with custom noise. Our results suggest that the corresponding QEM learned in one parameter regime of the model successfully transfers to a different parameter regime. In particular, our findings indicate that it transfers between different phases of the model. We observe that adiabatic GREC produces a smaller error than zero noise extrapolation (ZNE). Furthermore, in general, adiabatic GREC can be more cost-efficient in terms of the total number of gates used for the simulations. We comment on approaches to further reduce the necessary quantum computational resources. We also outline extensions of the introduced adiabatic GREC QEM method.
\end{abstract}

\maketitle

%%%%% MAIN BODY

\section{Introduction}

Quantum computing allows us to perform calculations in regimes inaccessible by classical numerical approaches~\cite{e2020-100571-8, ab6311, PRXQuantum.5.037001, 1.430.0228}. For instance, standard Monte Carlo methods are inapplicable to lattice Quantum Chromodynamics (LQCD) in the presence of a chemical potential or a topological $\theta$-term~\cite{deForcrand:2010j3, NAGATA2022103991}: These are examples of the so-called sign problem. However, while fully fault-tolerant quantum computing may be accomplished in the near future~\cite{PhysRevA.52.R2493, PhysRevLett.77.793, PhysRevA.54.1098, CBO9780511976667}, present-day quantum hardware is affected by quantum errors~\cite{0034-4885/61/2/002, DiVincenzo_2000, nature08812, RevModPhys.86.153, q-2018-08-06-79}. This will also hold for pioneering hardware in the fault-tolerant era~\cite{PRXQuantum.3.010345, arXiv.2501.05694, arXiv.2501.09079}. Therefore, current noisy intermediate-scale quantum (NISQ) devices require quantum error mitigation (QEM) of the performed measurements~\cite{s41534-022-00618-z, PhysRevLett.131.210602, RevModPhys.95.045005, s41567-024-02536-7}. To this end, various QEM techniques were developed, including~\cite{q-2021-11-26-592, PhysRevResearch.3.033098, q-2023-06-06-1034, PRXQuantum.2.040330, arXiv.2102.02120, PhysRevA.105.042408, arXiv.2502.21159, 9259940, PhysRevA.105.032620, s41534-019-0233-0, QCE57702.2023.00102, PhysRevA.58.2733, PhysRevLett.119.180509, PhysRevX.8.031027, ACCESS.2020.3031607, PhysRevApplied.20.064027, PhysRevA.94.052325, PhysRevA.104.052607, PhysRevX.7.021050, s41586-023-06096-3}.

\subsection{Problem-partitioning}

Due to the differences between the quantum and the classical computational methods outlined above, numerical tasks, such as those affected by the sign problem in LQCD, can be partitioned into two regimes: First, the ``classical-and-quantum'' (CL-QU) regime accessible to both standard classical and noisy quantum computations. Second the ``only-quantum'' (only-QU) regime accessible solely to noisy quantum computations. A subset of QEM techniques, see for example~\cite{q-2021-11-26-592, PhysRevResearch.3.033098, q-2023-06-06-1034, PRXQuantum.2.040330, arXiv.2102.02120, PhysRevA.105.042408}, is based on the idea of problem-partitioning: The properties of QEM learned in the CL-QU regime are extrapolated to the only-QU regime.

One such QEM method that is relevant to the present work is global randomized error cancellation (GREC)~\cite{PhysRevA.105.042408}. The GREC method can be outlined as follows: Controllable noise is added to the original noisy quantum circuit in the form of random insertions of parametrized gates. An observable of interest is measured in the CL-QU regime using the randomized quantum circuits, as well as on a classical device. A parametrized combination of the noisy measurements is found that best approximates the classical measurement. A mitigated quantum measurement of the observable in the only-QU regime is found by applying the previously learned combination's parameters to the measurements performed in the only-QU regime that are obtained with the same randomized realizations of the quantum circuits. We review GREC in more detail in Sec.~\ref{subsec:prelim_GREC}.

There are (at least) two different tasks that can be addressed regarding a QEM technique that utilizes the idea of problem-partitioning. First, the performance of the QEM technique. Second, a particular case of the first, where the CL-QU and the only-QU regions are in different phases of the model. We expect the second task to be a more challenging test of a QEM technique applied to a quantum computational method than the first, as some approaches may suffer from effects induced by the phase transition (PT). In the present work we study the Schwinger model~\cite{PhysRev.128.2425}, where for demonstration purposes we consider a sub-region of the parameter space to focus on providing an example of a combination of a quantum computational method and a QEM technique that can be successfully applied to the second task. The specific quantum computational method~\cite{arXiv.2407.09224} that we use is based on the adiabatic theorem of~\cite{BF01343193} (see the overview in Sec.~\ref{subsec:prelim_adiab_evol}).

Throughout the work, we refer to the CL-QU regime as the ``learning'' region $\mathcal{L}$ and to the only-QU regime as the ``prediction'' region $\mathcal{P}$. We assume that the considered parameter sub-space of the $\mathcal{P}$ region is in sufficient proximity to the $\mathcal{L}$ region.

\subsection{Applicability}

Studying phase diagrams is a substantial part of understanding physical theories. Regions of PTs are therefore of particular significance. The GREC QEM scheme was developed to be used for the mitigation of the effects of nonanalyticities in the proximity of PTs. An assumption of GREC is that the corresponding QEM scheme does not depend on nonanalyticities. However, this assumption has not yet been tested. In fact, for an order parameter with a nonanalyticity at a first-order PT we expect the properties of a GREC-like QEM scheme to differ in the regions separated by the PT. Within the adiabatic GREC studied in the present work, the adiabatic evolution of a single eigenstate follows the line of the corresponding energy level. All lines of energy levels are analytic even at PTs. The PTs are encoded in the crossings of these lines. This allows the adiabatic GREC QEM scheme to avoid problems of nonanalyticity.

The example of the Schwinger model~\cite{PhysRev.128.2425} (see Sec.~\ref{subsec:prelim_Schwinger_model} for a brief overview) illustrates the applicability of our mitigation method to a similar adiabatic evolution task motivated by the sign problem. Namely, the problem where the CL-QU regime includes both vanishing and small non-zero values of the background electric field $l_0$, while the only-QU regime is represented by arbitrarily large non-zero values of $l_0$. See e.g.~\cite{ab6311, deForcrand:2010j3, NAGATA2022103991} for overviews of solutions to tasks involving small non-zero values of parameters that lead to the sign problem. In the adiabatic evolution task motivated by the sign problem, the CL-QU regime must contain non-zero values of $l_0$ so that the properties of the QEM learned in that regime can be smoothly extrapolated to the only-QU regime.

\subsection{Goals}

Our first goal is to develop a GREC-like QEM scheme adapted to the mitigation of measurements obtained from a noisy quantum device by the adiabatic method in~\cite{arXiv.2407.09224} applied to the lattice Schwinger model (\ref{eqn:H_Schwinger}). For the adiabatic evolution in~\cite{arXiv.2407.09224}, the time-dependent adiabatic parameter is $l_0$ while all other parameters are fixed. Our second goal is to investigate whether the QEM of the energy levels learned by the adiabatic GREC method in one sub-domain of $l_0$ transfers to a different sub-domain of $l_0$ for the lattice Schwinger model (\ref{eqn:H_Schwinger}). In particular, within the $(l_0,m/g)$-parameter region ($m/g$ denotes the lattice mass over coupling ratio) containing the first-order PT at a critical $l_0^*$, we would like to study the following: Does the adiabatic GREC QEM of the energy levels transfer from the learning region $\mathcal{L}=[l_0^{\text{min}},l_0^{\text{int}}]$ with $l_0^* \notin \mathcal{L}$ to the prediction region $\mathcal{P}=(l_0^{\text{int}},l_0^{\text{max}}] \ni l_0^*$?~\footnote{The superscripts ``min'', ``int'' and ``max'' denote the minimal, intermediate and maximal $l_0$ values of the considered $l_0$ domain, respectively.} From here on, we define $l_0^*$ as the critical $l_0$ for which there is a first-order PT for large $m/g$ or a minimal non-zero gap for small $m/g$. An additional question that we address is how the adiabatic GREC mitigation scheme compares to other QEM methods. We study this on the example of zero noise extrapolation (ZNE)~\cite{9259940, QCE57702.2023.00102, PhysRevX.7.021050, PhysRevLett.119.180509}.

To summarize, our main goals are to test the following two claims:
\begin{itemize}
    \item[1.] GREC can be used for QEM in adiabatic evolution.
    \item[2.] QEM of energy levels via 1.\ transfers across phases.
\end{itemize}

\section{Preliminaries}

\subsection{GREC} \label{subsec:prelim_GREC}

Within the GREC method~\cite{PhysRevA.105.042408}, QEM of the measurement of an observable $\hat{A}$ is performed as follows: The original noisy quantum circuit is supplemented with randomly inserted parametrized gates. These correspond to different adjustable realizations of noise for the same noise level. The observable $\hat{A}$ is then measured over the CL-QU-domain $p_{\text{CL-QU}}$ of a parameter of the model $p$ for each randomized circuit realization $r=1, \dotsc, R$, as well as classically. Subsequently, a linear least squares (LLS) optimization problem that minimizes
\begin{equation} \label{eqn:GREC_minim}
\dfrac{1}{2} \sum\limits_{p \in p_{\text{CL-QU}}} \Bigg[ \left( \sum\limits_{r=1}^{R} \eta_r \braket{\hat{A}(p)}_{r}^{\text{noisy}} \right) + \eta_0 - \braket{\hat{A}(p)}^{\text{ideal}} \Bigg]^2
\end{equation}
is solved for $\eta_0, \eta_1, \dotsc, \eta_R$ to find a linear combination of noisy measurements $\braket{\hat{A}(p)}_{r}^{\text{noisy}}$ that best approximates the ideal classically-computed expectation values $\braket{\hat{A}(p)}^{\text{ideal}}$ in the CL-QU parameter domain. Here, the index $r$ in $\braket{\hat{A}(p)}_{r}^{\text{noisy}}$ labels the realizations of parametrized gates inserted into the original circuit. Each such realization is paired with a corresponding linear coefficient $\eta_r$. The coefficient $\eta_0$ is not associated with an individual noise realization. Note that in (\ref{eqn:GREC_minim}), the number of points considered from the $p_{\text{CL-QU}}$ domain must be greater than or equal to $R+1$ to ensure that the corresponding linear system is not underdetermined. The mitigated expectation value of the observable for $p \in p_{\text{only-QU}}$ is then obtained by
\begin{equation}
    \braket{\hat{A}(p)}^{\text{mitig}} = \left( \sum\limits_{r=1}^{R} \eta_r \braket{\hat{A}(p)}_{r}^{\text{noisy}} \right) + \eta_0 \, .
\end{equation}
In the present work, we extend GREC to QEM for adiabatic preparation of states. We develop our method in application to the finite-lattice Hamiltonian formulation~\cite{s41534-024-00950-6} of the Schwinger model~\cite{PhysRev.128.2425} in the context of~\cite{arXiv.2407.09224}.

\subsection{Adiabatic evolution} \label{subsec:prelim_adiab_evol}

The adiabatic evolution method of~\cite{arXiv.2407.09224} is based on the adiabatic theorem without a gap condition~\cite{BF01343193}. By this theorem, for an eigenstate $\ket{E_{\alpha}}$, corresponding to the energy level $E_{\alpha}$, prepared at time $t=0$ and adiabatically evolved for a time $T$, the transition probability to a different level $E_{\beta}$, $\alpha \neq \beta$, is given by
\begin{equation} \label{eqn:p_tr_adiab_thm}
    p_{\text{tr}} = \mathcal{O} \left( \left( e T \right)^{-2/(s+1)} \right) \, .
\end{equation}
Here, $e$ is a unit of energy (for the remainder of the work we set $e=1$, we also set $\hbar = c = 1$ throughout) and $s$ is the maximal order of the zeros of the function $E_{\alpha}(t/T)-E_{\beta}(t/T)$ for arbitrary indices $\alpha$, $\beta$ with $\alpha \neq \beta$. That is, $s$ is the measure of smoothness of the level-crossing: The smoother the crossing, the slower must the adiabatic evolution be for the evolved state to remain mainly on the same energy level. Nevertheless, with a sufficiently long evolution time, an adiabatically evolved eigenstate remains predominantly on its initial energy level, even if that level intersects with others. In~\cite{arXiv.2407.09224}, this enabled the recovery of the phase structure of the lattice Schwinger model via adiabatic evolution of the individual lowest-energy eigenstates without \emph{a priori} knowledge of the phase diagram, both in the no-phase-transition and the first-order-phase-transition phase diagram regions: The presence (absence) of a crossing of the lowest energy level with a higher energy level indicates the presence (absence) of a PT.

As the adiabatic GREC method is based on the adiabatic theorem, (\ref{eqn:p_tr_adiab_thm}) poses an applicability bound from the perspective of adiabaticity. In the context of the problem that we study, this bound is relevant to the size of the $l_0$-domain covered during the adiabatic evolution.

\subsection{The Schwinger model} \label{subsec:prelim_Schwinger_model}

Due to its relative simplicity and its similarity to QCD, the Schwinger model is a popular benchmark theory for novel lattice methods intended for future applications to LQCD, for example in quantum computing, see e.g.~\cite{nature18318, 2058-9565/ac5f5a, PRXQuantum.3.020324, arXiv.2401.08044, PRXQuantum.5.020315, PhysRevA.98.032331, q-2020-08-10-306, s41534-024-00950-6, PhysRevA.90.042305, Yamamoto:2022Qn, PhysRevD.105.094503, arXiv.2404.14788, s41586-019-1177-4, arXiv.2411.01079, arXiv.2409.13510, PhysRevD.106.054508, ptac007, PhysRevResearch.2.013288, PhysRevResearch.2.023342}.

The Schwinger model~\cite{PhysRev.128.2425} is a $U(1)$ gauge theory in (1+1)-dimensions coupled to a single flavor of a massive Dirac fermion. It can be discretized by the Kogut-Susskind approach~\cite{PhysRevD.11.395} and mapped to qubits by the Jordan-Wigner transformation~\cite{Jordan1928}. We use the specific formulation~\cite{s41534-024-00950-6} with open boundary conditions. The corresponding (rescaled) dimensionless Hamiltonian reads~\cite{s41534-024-00950-6}
\begin{equation} \label{eqn:H_Schwinger}
    \begin{aligned}
        H &= \dfrac{x}{2} \sum\limits_{n=0}^{N-2} \left( X_n X_{n+1} + Y_n Y_{n+1} \right) \\
        &+ \dfrac{1}{2} \sum\limits_{n=0}^{N-2} \sum\limits_{k=n+1}^{N-1} \left( N-k-1+\lambda \right) Z_n Z_k \\
        &+ \sum\limits_{n=0}^{N-2} \left( \dfrac{N}{4} - \dfrac{1}{2} \Bigl\lceil \dfrac{n}{2} \Bigr\rceil + l_0 \left( N-n-1 \right) \right) Z_n \\
        &+ \dfrac{m}{g} \sqrt{x} \sum\limits_{n=0}^{N-1} (-1)^n Z_n \\
        &+ l_0^2 \left( N-1 \right) + \dfrac{1}{2} l_0 N + \dfrac{1}{8} N^2 + \dfrac{\lambda}{4} N \, .
    \end{aligned}
\end{equation}
Here $X_n$, $Y_n$ and $Z_n$ are the Pauli operators acting on spin $n$, $N$ is an even number of sites, $x = (N/V)^2$ with dimensionless lattice volume $V$, $m/g$ is the lattice mass over coupling ratio, $l_0$ is the background electric field and $\lambda \gg 1$ is a Lagrange multiplier that enforces vanishing total charge.

Resulting from a topological $\theta$-term, a non-zero $l_0$ gives rise to the sign problem in standard classical numerical computations. In the continuum limit, at $|\text{frac}(l_0)|=1/2$ the Schwinger model exhibits a second-order PT point at the mass-over-coupling ratio of approximately $0.33$, a first-order PT line for higher values of the ratio, and no PTs for lower values, see e.g.~\cite{COLEMAN1975267, COLEMAN1976239, HAMER1982413, PhysRevD.66.013002, PhysRevD.97.014507, PhysRevD.101.054507, PhysRevD.105.014504, Pederiva:2022br, PhysRevD.108.014516, s41534-024-00950-6, PhysRevD.95.094509}.~\footnote{The continuum Schwinger model is periodic in $l_0=\theta/(2\pi)$ with a period of $1$~\cite{COLEMAN1976239}. The fractional part of $l_0$ is defined as $\text{frac}(l_0)=l_0-\lfloor | l_0 | \rfloor \cdot \text{sgn}(l_0)$.} The finite-size lattice Schwinger model has the same features, although the critical values of the parameters are shifted by finite-size effects of finite $N$ and $V$, such as an additive mass renormalization, see e.g.~\cite{PhysRevResearch.4.043133, PhysRevD.108.014516}.

\section{Method}

\subsection{Key idea}

Our time-dependent GREC QEM method is based on the following two observations: First, the $\eta$-coefficients in the original GREC algorithm~\cite{PhysRevA.105.042408} are constants. Second, the error of the time evolution on a noisy device accumulates with each additional time step. Consequently, the $\eta$-s need to be learned separately for each time point. This is the key idea of our mitigation scheme.

\subsection{Approach}

In the context of~\cite{arXiv.2407.09224}, our method aims to mitigate the noisy measurements of the energy of an individual instantaneous eigenstate at $l_0^{\text{min}}$ that is adiabatically evolved from $l_0(t=0)=l_0^{\text{min}}$ to $l_0(t=T)=l_0^{\text{max}}$. These energy measurements produce the corresponding energy level as a function of $l_0 \in \mathcal{L} \cup \mathcal{P} = [l_0^{\text{min}},l_0^{\text{max}}]$. To generate different sets of training data in the $\mathcal{L}$-region we perform the adiabatic evolution of the same initial instantaneous eigenstate over the same number of time steps $n_{\delta t}$, but with different $l_0 \in \mathcal{L}$ at the end of the evolution. Note that, in general, the values of $T$, $l_0(t=0)$ and $l_0(t=T)$ are permitted to differ among different training lines. See Fig.~\ref{fig:GRECmtg} for an example with the same $T$ and $l_0(t=0)$, and different $l_0(t=T)$ (and hence different $l_0$ step sizes $\delta l_0$).

\subsection{Procedure}

For clarity, in the following we will refer to the energy levels equivalently as lines. Each line adiabatically computed over $l_0 \in \mathcal{L} \cup \mathcal{P}$ with an added noise (AN) realization $r$ requires a corresponding set of training lines adiabatically computed over $l_0 \in \mathcal{L}$ with the same AN $r$. We label these training lines with an additional index $\tau=1,\dotsc,n_{\text{tr.lin.}}$. Given a line index $\alpha$, for each time point $i=0,\dotsc,n_{\delta t}$ we independently minimize
\begin{equation} \label{eqn:adiab_GREC_LLS}
\begin{aligned}
    \frac{1}{2} \sum\limits_{\tau=1}^{n_{\text{tr.lin.}}} \Bigg[ &\left( \sum\limits_{r=1}^{R_{\tau}} \eta_{\alpha,i,r} \left[ E_{\alpha}(l_{0,\tau}(i \, T_{\tau}/n_{\delta t})) \right]_{r}^{\text{noisy}} \right) \\
    &+ \eta_{\alpha,i,0} - \left[ E_{\alpha}(l_{0,\tau}(i \, T_{\tau}/n_{\delta t}))\right]^{\text{ideal}} \Bigg]^2
\end{aligned}
\end{equation}
to solve for the $\eta$-s in the $\mathcal{L}$-domain. For every training line $\tau$ in (\ref{eqn:adiab_GREC_LLS}), $R_{\tau}$ is the number of AN realizations, $l_{0,\tau}(t)$ is the adiabatic ramp and $T_{\tau}$ is the corresponding total adiabatic evolution time. Note that the number of time steps $n_{\delta t}$ is fixed universally for the to-be-mitigated line and for all of the training lines. The noisy data $\left[ E_{\alpha}(l_{0,\tau}(i \, T_{\tau}/n_{\delta t})) \right]_{r}^{\text{noisy}}$ corresponding to the AN realization $r$ of the training line $\tau$ is obtained by the method~\cite{arXiv.2407.09224} on a noisy quantum device, while $\left[ E_{\alpha}(l_{0,\tau}(i \, T_{\tau}/n_{\delta t}))\right]^{\text{ideal}}$ is computed classically. Using the $\eta$-s learned in the $\mathcal{L}$-domain, the mitigated values of the line over $l_0 \in \mathcal{L} \cup \mathcal{P}$ are obtained for each time point via
\begin{equation} \label{eqn:adiab_GREC_mitig}
\begin{aligned}
    &\left[ E_{\alpha}(l_0(i \, T/n_{\delta t})) \right]^{\text{mitig}} = \\
    &\left( \sum\limits_{r=1}^{R} \eta_{\alpha,i,r} \left[ E_{\alpha}(l_{0}(i \, T/n_{\delta t})) \right]_{r}^{\text{noisy}} \right) + \eta_{\alpha,i,0} \, ,
\end{aligned}
\end{equation}
where, from here on, we set $R = \max\limits_{1 \leq \tau \leq n_{\text{tr.lin.}}} R_{\tau}$ and take the unused coefficients $\eta_{\alpha,i,r}$ equal to zero.

\subsection{Illustration}

\begin{figure}[htbp]
    \centering
    \includegraphics[width=0.48\textwidth]{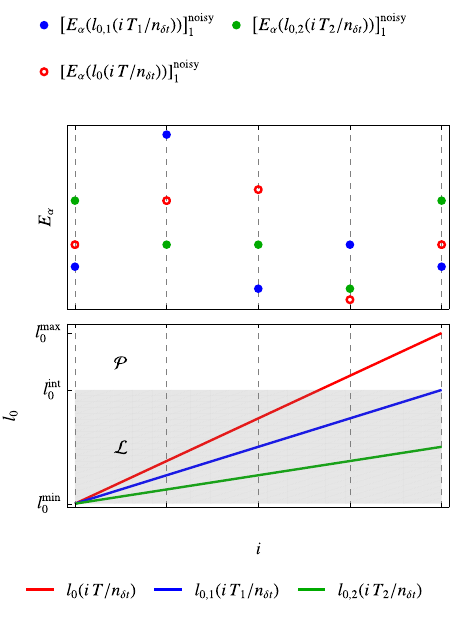}
    \caption{Schematic representation of the adiabatic GREC method. The data of the to-be-mitigated line is shown in red. The data of the training lines $\tau=1$ and $\tau=2$ is shown in blue and green, respectively. The bottom panel shows the adiabatic ramps over the time point index $i$. The top panel shows the corresponding noisy energy measurements for the line index $\alpha$ over the time point index $i$. For a complementary plot of $E_{\alpha}$ over $l_0$ see the results shown in Fig.~\ref{fig:GRECmtg}.}
    \label{fig:GRECschm}
\end{figure}

Figure~\ref{fig:GRECschm} shows a schematic representation of the adiabatic GREC method for a single line index $\alpha$. For clarity, Fig.~\ref{fig:GRECschm} displays the basic necessary set-up, i.e.\ $R = 1$ and $n_{\text{tr.lin.}}=2$. For a single line index $\alpha$, at each time point $i$, the training data in $\mathcal{L}$, namely $\left[ E_{\alpha}(l_{0,\tau}(i \, T_{\tau}/n_{\delta t})) \right]_{1}^{\text{noisy}}$ for $\tau=1,2$ (blue and green points in the top panel of Fig.~\ref{fig:GRECschm}) and $\left[ E_{\alpha}(l_{0,\tau}(i \, T_{\tau}/n_{\delta t})) \right]^{\text{ideal}}$, is used in the LLS (\ref{eqn:adiab_GREC_LLS}) to solve for $\eta_{\alpha,i,0}$ and $\eta_{\alpha,i,1}$. The to-be-mitigated data $\left[ E_{\alpha}(l_{0}(i \, T/n_{\delta t})) \right]_{1}^{\text{noisy}}$ in $\mathcal{L} \cup \mathcal{P}$ (red ``$\circ$''-s in the top panel of Fig.~\ref{fig:GRECschm}), along with the obtained values of the coefficients $\eta_{\alpha,i,0}$ and $\eta_{\alpha,i,1}$, is then used in (\ref{eqn:adiab_GREC_mitig}) to find the mitigated data $\left[ E_{\alpha}(l_0(i \, T/n_{\delta t})) \right]^{\text{mitig}}$ in $\mathcal{L} \cup \mathcal{P}$.

\subsection{Hyperparameter selection}

The adiabatic GREC method can be extended to allow for hyperparameter selection. Similarly to the original GREC, the region with the training data could be partitioned into the learning region $\mathcal{L}$ and a ``testing'' region $\mathcal{T}$. For example, one part of the training data would be used in $\mathcal{L}$ for the minimization task to learn the $\eta$-s of distinct combinations of training lines, and the other part would be used in $\mathcal{T}$ to select the optimal of the learned subsets of $\eta$-s. Here, the selection among the combinations of training lines represents the hyperparameter optimization. Furthermore, the $\mathcal{T}$ region could be used to choose among realizations of different noise levels. Here, the level of the added noise plays the role of the hyperparameter.

In our simulations, we take $\mathcal{T}=\emptyset$ and consider a single noise level for simplicity. For the numerical analysis in Sec.~\ref{sec:results} we perform an \emph{a posteriori} hyperparameter selection:
Here, the hyperparameter is the number of training lines $n_{\text{tr.lin.}}$.
We study how the error of the mitigation method in the $\mathcal{P}$ region depends on $n_{\text{tr.lin.}}$. The results are shown in Fig.~\ref{fig:cfPregRZ}.

\section{Numerical framework}

\subsection{Adiabatic evolution and lattice parameters}

We perform a Lie-Trotter expansion~\cite{2033649, s00220-006-0150-x, 11526216_2, cpa.3160070404, BF01609348} of the time evolution operator and take $n_{\delta t}=10^2$. For the to-be-mitigated line over $l_0 \in \mathcal{L} \cup \mathcal{P}$ we use the linear ramp
\begin{equation} \label{eqn:l_0_ramp}
    l_0(t)=l_0^{\text{min}}+ \left( l_0^{\text{max}}-l_0^{\text{min}} \right) \dfrac{t}{T} \, .
\end{equation}
We perform simulations for $n_{\text{tr.lin.}}=2,\dotsc,11$ training lines, for which we use analogous linear ramps
\begin{equation} \label{eqn:l_0_tau_ramps}
    l_{0,\tau}(t)=l_0^{\text{min}}+ \left( l_0^{\text{int}}-l_0^{\text{min}} \right) \dfrac{\left( 11-\tau \right)}{10} \dfrac{t}{T_{\tau}} \, .
\end{equation}
For the evolution times we set $T_{\tau}=T \, \forall \tau$ and $T=10$. The remaining parameter values for (\ref{eqn:H_Schwinger}) are $N=6$, $V=N/\sqrt{x}=30$ and $\lambda=100$, and we consider the two regimes $m/g \in \{0,10\}$ ($\{$no level crossing, a level crossing$\}$ between the two lowest energy levels over $l_0 \in \mathcal{L} \cup \mathcal{P}$, respectively). Here, for demonstration purposes, we restrict our simulations to small $l_0$-domains (see Tab.~\ref{tab:l_0_domains}). Following~\cite{arXiv.2407.09224}, we consider the two lowest energy levels $E_{\alpha=0,1}$. The indices $\alpha=0$ and $\alpha=1$ denote the ground state at $l_0^{\text{min}}$ and the first-excited state at $l_0^{\text{min}}$, respectively.

\begin{table}[htbp]
\centering
\begin{tabular}{l | l l l l}
$m/g$ & $l_0^{\text{min}}$ & $l_0^{\text{int}}$ & $l_0^*$ & $l_0^{\text{max}}$ \\
\hline
$0$ & $0.511527$ & $0.512187$ & $0.512360$ & $0.512527$ \\
$10$ & $1.832633$ & $1.833293$ & $1.833466$ & $1.833633$
\end{tabular}
\caption{Values of $l_0^{\text{min}}$, $l_0^{\text{int}}$, $l_0^*$ and $l_0^{\text{max}}$ for $m/g \in \{0,10\}$.}
\label{tab:l_0_domains}
\end{table}

\subsection{Intrinsic noise model}

We perform the adiabatic evolution on a digital quantum simulator with the gate basis $\{$CX, ID, RZ, SX, X$\}$ using Qiskit~\cite{arXiv:2405.08810}. Here, for demonstration purposes, the initial state is prepared by exact diagonalization (ED). We compute the ``ideal'' values by ED over the entire $l_0$ domain with QuSpin~\cite{SciPostPhys.2.1.003, SciPostPhys.7.2.020, QuSpin_doc}. The quantum simulations are performed with $n_{\text{shots}}=10^4$ shots. Due to the large depth of the circuits, we found both our method and ZNE applicable to the noisy data only for a custom mild-noise model. Specifically, we consider a noise model where a $Z$-flip in the RZ gate has a probability of $10^{-6}$. We have verified that for the measured energy values the combined error due to the Trotter expansion, the shot noise and the non-adiabaticity of the evolution (due to finite $T$) is negligible compared to the error due to the noise model.

\subsection{Added noise}

We implement the minimal version of our method: We consider a single AN realization, i.e.\ $R = 1$. Our goal is to mitigate the expectation value of a time-evolved observable. Therefore, we choose to add noise only to the Hamiltonian and not to the initial state $\ket{E_{\alpha}}$ at $l_0^{\text{min}}$. This parametrized noise can be added at the level of the Hamiltonian or equivalently directly at the circuit level. For simplicity, we choose the former option. To preserve the Hermiticity of the Hamiltonian, each Pauli operator $P \in \{ X, Y, Z\}$ in (\ref{eqn:H_Schwinger}) is mapped as $P \to W_P^{\dagger} P W_P$ with $W_P$ a complex $2 \times 2$ matrix. For simplicity, we restrict $W_P$ to the rotation
\begin{equation}
    W_P =
    \begin{pmatrix}
        \cos (\phi^P_1/2) & -\mathrm{e}^{i \phi^P_3} \sin (\phi^P_1/2) \\
        \mathrm{e}^{i \phi^P_2} \sin (\phi^P_1/2) & \mathrm{e}^{i (\phi^P_2+\phi^P_3)} \cos (\phi^P_1/2) \\
    \end{pmatrix}
\end{equation}
with $\phi^P_1,\phi^P_2=0$ and sample $\phi^P_3$ uniformly at random on the interval $[-1,1]$. For $R=1$, a single line over $l_0 \in \mathcal{L} \cup \mathcal{P}$ is parametrized by $\eta$-s. Therefore, the total number of simulations where we evolve an energy level is $n_{\text{evol}}^{\text{GREC}}=n_{\text{tr.lin.}}+1 \geq 3$ (see Fig.~\ref{fig:GRECschm} for a schematic representation and Fig.~\ref{fig:GRECmtg} for an example). We apply our mitigation scheme for $n_{\text{evol}}^{\text{GREC}}=3,\dotsc,12$.

\subsection{ZNE for adiabatic evolution}

To evaluate the efficiency of our method, we also perform ZNE of the line over $l_0 \in \mathcal{L} \cup \mathcal{P}$. Here, no parametrized gates are inserted into the original circuit. Instead, there are local unitary foldings of the form $(U^{\dagger}U)^{(f-1)/2}$ after each circuit $U$ representing the time evolution for a single Trotter slice, where $f=2q-1$ is the noise factor, $q=1,\dotsc,n_{\text{evol}}^{\text{ZNE}}$ and $n_{\text{evol}}^{\text{ZNE}} \geq 2$ is the total number of simulations for a given energy level. Each Trotter slice corresponds to a time evolution of the system at a different value of $l_0$. We therefore choose to perform the unitary foldings locally, after each individual Trotter slice. This leads to the quantum computational cost of ZNE discussed in Sec.~\ref{subsubsec:quant_comp_cost}.

For the time evolution, we perform ZNE for each individual time point $i=0,\dotsc,n_{\delta t}$ (for an example of ZNE QEM results see Fig.~\ref{fig:ZNE}). We apply the ZNE for $n_{\text{evol}}^{\text{ZNE}}=2,\dotsc,10$. We find a linear fit to be appropriate for the obtained data.

\subsection{Performance metrics}

\subsubsection{Quantum computational cost} \label{subsubsec:quant_comp_cost}

We would now like to compare the quantum computational costs of ZNE and adiabatic GREC. One way to do this is to calculate the total number of applications of gates G $\in \{$CX, RZ, SX$\}$ required for all of the simulations. The number of gates G $\in \{$CX, RZ, SX$\}$ that form a single Trotter slice (TS) circuit for both the original circuit (ORIG) and GREC is given in Tab.~\ref{tab:n_G_per_TS}.

\begin{table}[htbp]
\centering
\begin{tabular}{c c | c c c}
  &   & \multicolumn{3}{c}{$n_{\text{G}/\text{TS}}$} \\
\cline{2-5}
  & G & CX & RZ & SX \\
\hline
\multicolumn{1}{ l }{\multirow{2}{*}{ORIG} }        &
\multicolumn{1}{|l|}{$m/g=0$ } & $50$ & $130$ & $40$ \\
\multicolumn{1}{ c }{}                             &
\multicolumn{1}{|l|}{$m/g=10$} & $50$ & $131$ & $40$ \\
\hline
\multicolumn{1}{ l }{\multirow{2}{*}{GREC} }       &
\multicolumn{1}{|l|}{$m/g=0$ } & $70$ & $240$ & $80$ \\
\multicolumn{1}{ c }{}                             &
\multicolumn{1}{|l|}{$m/g=10$} & $70$ & $241$ & $80$
\end{tabular}
\caption{Number of gates per Trotter slice $n_{\text{G}/\text{TS}}$ with gate $\text{G} \in \{\text{CX}, \text{RZ}, \text{SX}\}$, for the original circuit (ORIG) and for the adiabatic GREC circuit with parametrized added noise, for $m/g \in \{0,10\}$. Note that for $m/g \neq 0$ there is a $Z_n$ acting on the spin $n=N-1$ (see (\ref{eqn:H_Schwinger})), resulting in an additional RZ gate.}
\label{tab:n_G_per_TS}
\end{table}

The number of gates required to prepare the initial state is the same for ZNE and adiabatic GREC. Therefore, we only compare the total number of gates $N_{\text{G}}^{\text{EM}}$ necessary for the respective time evolutions for $\text{EM} \in \{\text{ZNE},\text{GREC}\}$. These are
\begin{equation} \label{eqn:N_G_EM}
    \begin{aligned}
        N_{\text{G}}^{\text{ZNE}} &= |S| \dfrac{n_{\text{G}/\text{TS}}^{\text{ORIG}}}{2} (n_{\delta t}+1) (n_{\delta t}+2) \left( n_{\text{evol}}^{\text{ZNE}} \right)^2 \, , \\
        N_{\text{G}}^{\text{GREC}} &= |S| \dfrac{n_{\text{G}/\text{TS}}^{\text{GREC}}}{2} (n_{\delta t}+1) (n_{\delta t}+2) n_{\text{evol}}^{\text{GREC}} \, ,
    \end{aligned}
\end{equation}
where we take $R=1$ for adiabatic GREC. In (\ref{eqn:N_G_EM}), $S$ is a subset of energy levels. Note from (\ref{eqn:N_G_EM}) that while $N_{\text{G}}^{\text{GREC}}$ is linear in $n_{\text{evol}}^{\text{GREC}}$, $N_{\text{G}}^{\text{ZNE}}$ is quadratic in $n_{\text{evol}}^{\text{ZNE}}$ due to additional local unitary foldings at each time point for every extra line evolved.

\subsubsection{Total error}

Since the ideal data over $\mathcal{L}$ is used to learn the $\eta$-s, we would like to compare the errors of ZNE and adiabatic GREC only over $\mathcal{P}$. For $\text{EM} \in \{\text{ZNE},\text{GREC}\}$ we define the error in $\mathcal{P}$ as
\begin{equation} \label{eqn:error_in_P}
\begin{aligned}
    \mathcal{E}^{\text{EM}} = \sum\limits_{\alpha \in S} \Bigg( \dfrac{1}{|I|} \sum\limits_{i \in I} \big[ &E_{\alpha}^{\text{ED}}(l_0(i \, T/n_{\delta t})) \\
    - &E_{\alpha}^{\text{EM}}(l_0(i \, T/n_{\delta t})) \big]^2 \Bigg)^{1/2} \, ,
\end{aligned}
\end{equation}
where $S$ is a subset of energy levels and $I$ is the subset of time points $i$ such that $l_0(i \, T/n_{\delta t}) \in \mathcal{P}$. We denote by $\mathcal{E}^{\text{noisy}}$ the analogous error in $\mathcal{P}$ of the unmitigated lines $\alpha \in S$ evolved by the original noisy circuit. We consider $S=\{0,1\}$.

We find the error from the added parametrized noise in adiabatic GREC comparable to that from a single folding $U^{\dagger} U$ at each time point in ZNE (see AN in Fig.~\ref{fig:GRECres} and $f=3$ in Fig.~\ref{fig:ZNE}). We present our results below.

\section{Results} \label{sec:results}

\subsection{Adiabatic GREC}

\begin{figure*}[htbp]
    \centering
    \includegraphics[width=1.0\textwidth]{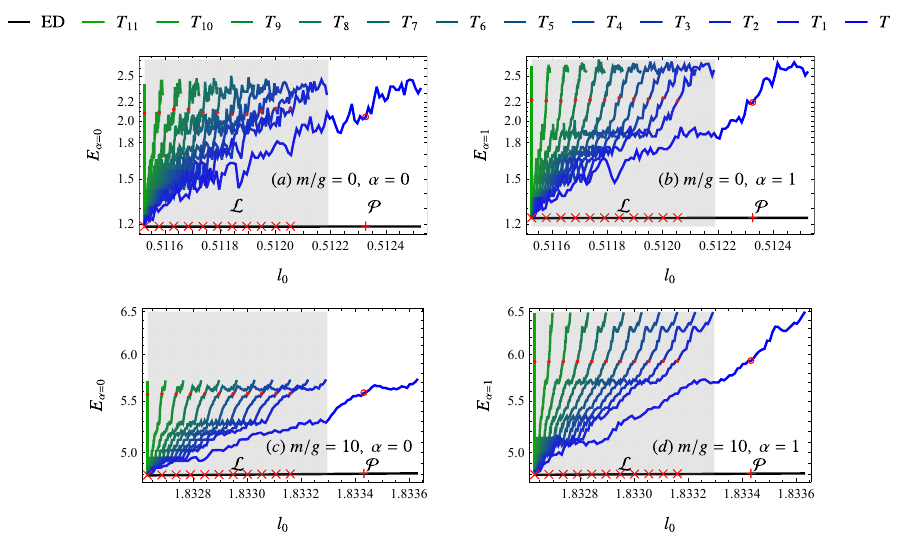}
    \caption{Added noise lines for adiabatic GREC. Panels (a), (b) and (c), (d) show the added noise lines for $m/g=0$ and $m/g=10$, respectively. The indices $\alpha=\{0,1\}$ are assigned to the levels corresponding to the $\{$ground, first-excited$\}$ states at $l_0^{\text{min}}$, respectively. The to-be-mitigated line over $l_0 \in \mathcal{L} \cup \mathcal{P}$ is labeled as $T$. The corresponding training lines $\tau=1,\dotsc,11$ over $l_0 \in \mathcal{L}$ are labeled as $T_{\tau}$. The ``ideal'' (solid black) line is labeled as ED. The red points and the red ``$\circ$'' mark the exemplary $i=80$ time point for the training and the to-be-mitigated lines, respectively. The red ``$\times$''-s and the red ``$+$'' mark the points on the ED line for the corresponding $l_0$.}
    \label{fig:GRECmtg}
\end{figure*}

Figure~\ref{fig:GRECmtg} shows the AN realizations of both the line over $l_0 \in \mathcal{L} \cup \mathcal{P}$ and the training lines over $l_0 \in \mathcal{L}$. Note that for the learning runs with different numbers of training lines $n_{\text{tr.lin.}}=2,\dotsc,11$ we select the $n_{\text{tr.lin.}}$ lines with $l_{0,\tau}(T)$ nearest to the $\mathcal{P}$-region to maximize the proximity of the training data to the to-be-mitigated line (see Fig.~\ref{fig:GRECmtg}, (\ref{eqn:l_0_ramp}) and (\ref{eqn:l_0_tau_ramps})).

We remark that in Fig.~\ref{fig:GRECmtg} the ED lines from panels (a) and (b) exhibit a minimal non-zero gap at $l_0^*$, while the ED lines from panels (c) and (d) cross at $l_0^*$. See Tab.~\ref{tab:l_0_domains} for the corresponding values of $l_0^*$. At the scale chosen in Fig.~\ref{fig:GRECmtg}, these features are not visible. We refer to the figures of~\cite{arXiv.2407.09224} for a clear presentation of these features.

\begin{figure*}[htbp]
    \centering
    \includegraphics[width=0.85\textwidth]{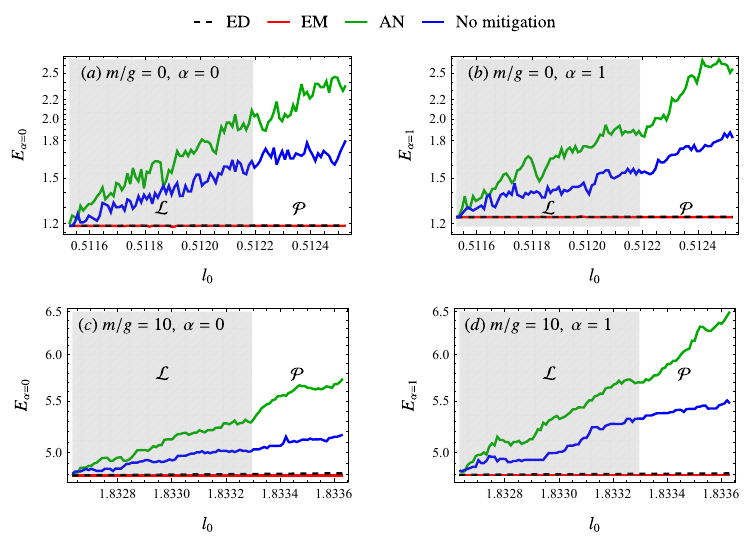}
    \caption{Results of QEM by adiabatic GREC. Panels (a), (b) and (c), (d) show the results for $\mathcal{E}^{\text{GREC}}_{\text{min}}$ in $\mathcal{P}$ (see (\ref{eqn:error_in_P})) for $m/g=0$ ($n_{\text{evol}}^{\text{GREC}}=7$) and $m/g=10$ ($n_{\text{evol}}^{\text{GREC}}=12$), respectively. The indices $\alpha=\{0,1\}$ are assigned to the levels corresponding to the $\{$ground, first-excited$\}$ states at $l_0^{\text{min}}$, respectively. The to-be-mitigated line with added noise (AN) is shown in solid green. The corresponding error mitigated line (EM) and the ``ideal'' (ED) line are shown in solid red and dashed black, respectively. For reference, the unmitigated line (No mitigation) computed via the original noisy circuit is shown in solid blue.}
    \label{fig:GRECres}
\end{figure*}

Figure~\ref{fig:GRECres} shows the best-obtained results of QEM by adiabatic GREC. We observe that the ``ideal'' line (ED) and the error-mitigated line (EM) are mostly indistinguishable at this scale. We observe from Fig.~\ref{fig:GRECres} that the $\eta$-s learned in the $\mathcal{L}$-region can be successfully applied in the $\mathcal{P}$-region for QEM of the two lowest energy levels in both of the following cases: First, where there is no crossing between the levels and correspondingly no PT ($m/g=0$). Second, where there is a level crossing and a corresponding PT ($m/g=10$).

With the $\left[E_{\alpha}\left(l_0\right)\right]^{\text{ideal}}$ ED data from the $\mathcal{L}$-region at hand, we can study how the noise depends on $l_0$. We observe no such significant dependence for $m/g \in \{0,10\}$, $\alpha \in \{0,1\}$. Moreover, for adiabatic GREC, the mitigated line $\left[E_{\alpha}\left(l_0\right)\right]^{\text{mitig}}$ can be additionally adjusted for a trend in $l_0$ using the data in $\mathcal{L}$ after the minimizations at individual $i=0,\dotsc,n_{\delta t}$. Our data for both $m/g=0$ and $m/g=10$ indicates a negligible linear trend $\left[E_{\alpha}\left(l_0\right)\right]^{\text{ideal}} - \left[E_{\alpha}\left(l_0\right)\right]^{\text{mitig}} = a \cdot l_0 + b$ with $a,b>0$ (see Fig.~\ref{fig:GRECres}). We neglect this correction as already without it the error of QEM by adiabatic GREC is smaller than that by ZNE (see Tab.~\ref{tab:errors}). As an additional remark, we note that for both ZNE and adiabatic GREC we are not able to recover the $l_0^*$ of the minimal non-zero gap ($m/g=0$) and the PT ($m/g=10$) from the corresponding mitigated data due to noise.

\subsection{Comparison with ZNE}

\begin{figure*}[htbp]
    \centering
    \includegraphics[width=0.85\textwidth]{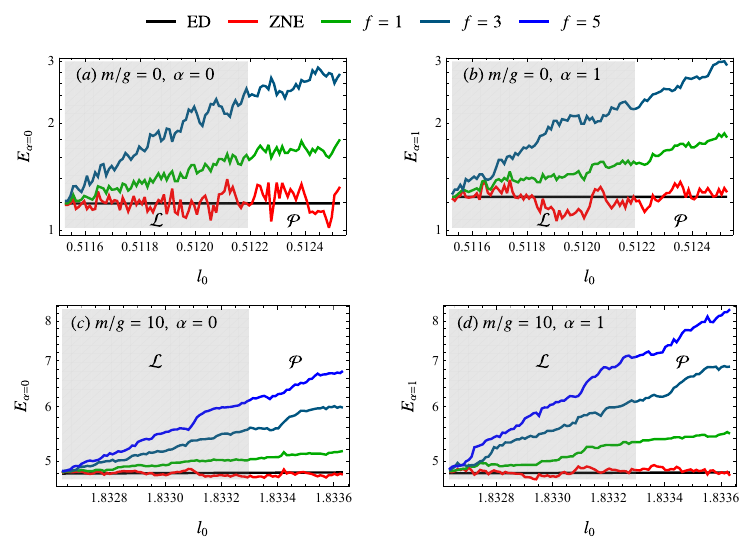}
    \caption{Example of ZNE QEM results. The panel labeling is the same as that in Fig.~\ref{fig:GRECmtg}. The panels show the lines associated with $\mathcal{E}^{\text{ZNE}}_{\text{min}}$ in $\mathcal{P}$ (see (\ref{eqn:error_in_P})) for $m/g=0$ ($f=1,3$) and $m/g=10$ ($f=1,3,5$). The ED and the ZNE lines are shown in solid black and solid red, respectively. The domains $\mathcal{L}$ and $\mathcal{P}$ are shown to aid the comparison of ZNE with adiabatic GREC.}
    \label{fig:ZNE}
\end{figure*}

Figure~\ref{fig:ZNE} shows the results of QEM by time-point-wise ZNE (similar to Fig.~\ref{fig:GRECres} for adiabatic GREC). The results indicate that the line mitigated by ZNE exhibits greater deviations from the ED line than the line mitigated by adiabatic GREC. We quantify this below.

Table~\ref{tab:errors} shows the errors (minimized over $n_{\text{evol}}^{\text{EM}}$ for $\text{EM} \in \{\text{ZNE},\text{GREC}\}$) as well as the improvement of relative error $(\mathcal{E}^{\text{GREC}}_{\text{min}}-\mathcal{E}^{\text{ZNE}}_{\text{min}})/\mathcal{E}^{\text{noisy}}$ in $\mathcal{P}$ for $m/g=\{0,10\}$. Our results suggest that adiabatic GREC produces a smaller error than ZNE for the considered problem. For the error over $\mathcal{L} \cup \mathcal{P}$ we obtain results analogous to those over $\mathcal{P}$.

\begin{table}[htbp]
\centering
\begin{tabular}{l | l l l | c}
$m/g$ & $\mathcal{E}^{\text{noisy}}$ & $\mathcal{E}^{\text{ZNE}}_{\text{min}}$ & $\mathcal{E}^{\text{GREC}}_{\text{min}}$ & $(\mathcal{E}^{\text{GREC}}_{\text{min}}-\mathcal{E}^{\text{ZNE}}_{\text{min}})/\mathcal{E}^{\text{noisy}}$ \\
\hline
$0$ & $0.966$ & $0.158$ & $\mathbf{0.002}$ & $-16.1\%$ \\
$10$ & $0.900$ & $0.102$ & $\mathbf{0.026}$ & $-8.5\%$
\end{tabular}
\caption{Errors $\mathcal{E}^{\text{noisy}}$, $\mathcal{E}^{\text{ZNE}}_{\text{min}}$, $\mathcal{E}^{\text{GREC}}_{\text{min}}$ and the improvement of relative error $(\mathcal{E}^{\text{GREC}}_{\text{min}}-\mathcal{E}^{\text{ZNE}}_{\text{min}})/\mathcal{E}^{\text{noisy}}$ in $\mathcal{P}$ for $m/g \in \{0,10\}$.}
\label{tab:errors}
\end{table}

\begin{figure*}[htbp]
    \centering
    \includegraphics[width=1.0\textwidth]{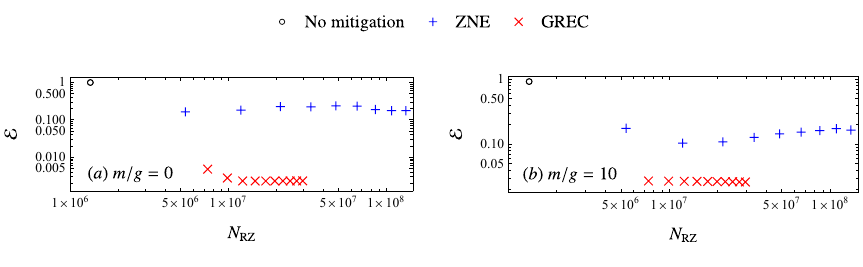}
    \caption{Comparison of ZNE and adiabatic GREC. The error $\mathcal{E}$ in $\mathcal{P}$ is shown over the total number of RZ gates $N_{\text{RZ}}$. The displayed errors are $\mathcal{E}^{\text{noisy}}$ computed from the noisy data without mitigation, $\mathcal{E}^{\text{ZNE}}$ and $\mathcal{E}^{\text{GREC}}$. The corresponding unmitigated, ZNE and GREC data is marked by a black ``$\circ$'', blue ``$+$''-s and red ``$\times$''-s, respectively. The ZNE and GREC data is shown for $N_{\text{RZ}}$ corresponding to $n_{\text{evol}}^{\text{ZNE}}=2, \dotsc, 10$ and $n_{\text{evol}}^{\text{GREC}}=3, \dotsc, 12$, respectively (see (\ref{eqn:N_G_EM})). The results for $m/g = 0$ and $m/g = 10$ are shown in panels (a) and (b), respectively.}
    \label{fig:cfPregRZ}
\end{figure*}

Figure~\ref{fig:cfPregRZ} shows plots of $\mathcal{E}^{\text{EM}}$ in $\mathcal{P}$ over $N_{\text{RZ}}^{\text{EM}}$ for $\text{EM} \in \{ \text{ZNE}, \text{GREC} \}$ for $m/g \in \{0,10\}$. The error $\mathcal{E}^{\text{noisy}}$ in $\mathcal{P}$ is also shown for reference. In Fig.~\ref{fig:cfPregRZ}, points at increasing $N_{\text{RZ}}$ correspond to increasing $n_{\text{evol}}^{\text{EM}}$, $\text{EM} \in \{ \text{ZNE}, \text{GREC} \}$ (see (\ref{eqn:N_G_EM})). We observe that, first, adiabatic GREC has a smaller error than ZNE for all considered simulations. Second, compared to a mitigation by ZNE with non-minimal resources ($n_{\text{evol}}^{\text{ZNE}} \geq 3$), a mitigation by adiabatic GREC can be more cost-efficient in terms of $N_{\text{RZ}}$. We obtain analogous results for the gates CX and SX.

\section{Conclusion}

\subsection{Summary}

In this work we have extended the GREC method~\cite{PhysRevA.105.042408} to QEM of adiabatically evolved states on a noisy quantum device. The original GREC method~\cite{PhysRevA.105.042408} is an example of a QEM technique where the problem is partitioned into two regimes: The CL-QU regime accessible to both standard classical and noisy quantum computations, and the only-QU regime accessible only to noisy quantum computations. For the error mitigation of the noisy quantum measurements, the features of QEM learned in the CL-QU regime are extrapolated to the only-QU regime.

We have developed the adiabatic GREC method in the context of our previous work~\cite{arXiv.2407.09224}. Based on the adiabatic theorem of~\cite{BF01343193}, in~\cite{arXiv.2407.09224} we have adiabatically evolved the lowest energy eigenstates to recover the phase structure of the lattice Schwinger model~\cite{s41534-024-00950-6} in the following two cases: In the presence (absence) of a level crossing with the corresponding presence (absence) of a PT.

In the present work, first, by formulating the adiabatic GREC method we have demonstrated that GREC can be used for QEM in adiabatic evolution of states. The key ingredient of our method is to perform GREC at each time point individually. Second, we have found evidence that QEM by adiabatic GREC of adiabatically evolved energy levels transfers across different parameter regions of the model. In particular, our results suggest that it transfers between different phases (i.e.\ across PTs). Furthermore, we have argued that adiabatic GREC can also be applied in the identical manner to an adiabatic evolution problem where the QEM characteristics learned in a CL-QU region are extrapolated to an only-QU region.

We have also found support for the following expectation: for an order parameter with a nonanalyticity at a first-order PT, the properties of a GREC-like QEM scheme differ in the regions separated by the PT. Specifically, the first-order derivative of the ground state energy of the system with respect to the model parameter that is relevant to the PT exhibits a discontinuity at a first-order PT. The lowest energy level corresponds to two different energy levels on the left and on the right of the PT. The $\eta$-coefficients used for QEM within the adiabatic GREC method are different for each non-degenerate energy level. Therefore, within adiabatic GREC, the $\eta$-coefficients of the mitigated lowest energy level will be different in the regions separated by the PT.

Due to the large depth of the time-evolution circuits, we have tested our mitigation method for a custom error model with custom noise. We have implemented the minimal version of the adiabatic GREC method by considering a single realization of added noise and by omitting the selection of optimal subsets of added noise realizations. We have compared our mitigation method to ZNE. Our results suggest that, first, adiabatic GREC produces a smaller error than ZNE by at least one order of magnitude. Second, except for the minimal-resources version of ZNE, adiabatic GREC can be more cost-efficient than ZNE in terms of the total number of gates used for the simulations. See Tab.~\ref{tab:errors} and Fig.~\ref{fig:cfPregRZ} for details.

\subsection{Outlook}

In the present implementation of adiabatic GREC the parametrized noise is added at the level of the Hamiltonian. As an outlook for future work, we instead propose adding noise by inserting parametrized gates directly at the circuit level. We expect that this will reduce the quantum computational cost in terms of the total number of gates used for the simulations without greatly reducing the efficiency of the adiabatic GREC QEM method. We leave the detailed analysis of how the error of the adiabatic GREC QEM method scales with the number of parametrized gate insertions applied directly at the circuit level for future work.

To conclude, we propose two further QEM methods in application to adiabatic time evolution, inspired by adiabatic GREC. Both of these use added noise in the form of parametrized gate insertions. The first method is a modification of ZNE. Here, the noise factor is governed by the number of parametrized gate insertions instead of the number of local unitary foldings. The magnitude of the added noise is the same for all inserted gates. Similarly to the original ZNE, this modified version of ZNE is performed for each individual time point. The second method is an extension of adiabatic GREC studied in the present work. Similarly to adiabatic GREC, each added noise realization parametrizes the noise for the corresponding set of training lines. However, here, each added noise realization has a different noise level. Just as in the adiabatic GREC, this extended adiabatic GREC is performed for each individual time point.

%%%%% ACKNOWLEDGMENTS & DISCLAIMERS

\begin{acknowledgments}

This work has received support from the French State managed by the National Research Agency under the France 2030 program with reference ANR-22-PNCQ-0002. We acknowledge the use of IBM Quantum services for this work.

The views expressed are those of the authors, and do not reflect the official policy or position of IBM or the IBM Quantum team.

\end{acknowledgments}

\section*{Data availability}

The data that support the findings of this article are available from the authors upon reasonable request.

% %%%%% BIBLIOGRAPHY
% 
% %%% custom style references
% \let\oldclearpage\clearpage
% \let\clearpage\relax
% \include{Custom_refs}

%%%%% APPENDIX

% \let\clearpage\oldclearpage
% \clearpage

% \setcounter{secnumdepth}{3}

% \begin{center}
%     {\large\textbf{Supplemental Material}}
% \end{center}

% \clearpage

% \begin{appendix}

% \end{appendix}

%%%%% BIBLIOGRAPHY

%%% custom style references
\let\clearpage\relax

\end{document}